\documentclass[twocolumn]{aastex631}

\usepackage{newtxtext,newtxmath}

\usepackage[T1]{fontenc}

\DeclareRobustCommand{\VAN}[3]{#2}
\let\VANthebibliography\thebibliography
\def\thebibliography{\DeclareRobustCommand{\VAN}[3]{##3}\VANthebibliography}

\usepackage{graphicx}	
\usepackage{amsmath}	
\usepackage{amssymb}	
\usepackage{color}
\usepackage{xcolor}

\newcommand{\Mpc}{\mathrm{~km~s^{-1}~Mpc^{-1}}}

\newcommand{\LCDM}{\rm{\Lambda}CDM}

\begin{document}

\title{Revisiting the Hubble constant, sound horizon and cosmography from late-time Universe observations}

\author{Zhiwei Yang}
\affiliation{College of Physics, Hebei Normal University, Shijiazhuang 050024, China}
\affiliation{Shijiazhuang Key Laboratory of Astronomy and Space Science, Hebei Normal University, Shijiazhuang 050024, China}
\author{Tonghua Liu$^\dagger$}
\affiliation{School of Physics and Optoelectronic, Yangtze University, Jingzhou 434023, China}
\email{$\dagger$ liutongh@yangtzeu.edu.cn}

\author{Xiaolei Li$^\star$}
\affiliation{College of Physics, Hebei Normal University, Shijiazhuang 050024, China}
\affiliation{Shijiazhuang Key Laboratory of Astronomy and Space Science, Hebei Normal University, Shijiazhuang 050024, China}
\email{$\star$ lixiaolei@hebtu.edu.cn}

\begin{abstract}

{The Hubble tension has become one of the central problems in cosmology. In this work, we determine the Hubble constant $H_0$ and sound horizon $r_d$ by using the combination of Baryon Acoustic Oscillations (BAOs) from DESI surveys, time-delay lensed quasars from H0LiCOW collaborations and the Pantheon supernovae observations. We consider two cosmological approaches, i.e., Taylor series and Pad\'{e}  polynomials, to avoid cosmological dependence. The reason for using this combination of data is that the absolute distance provided by strong gravitational lensing helps anchor the relative distance of BAO, and supernovae provide a robust history of universe evolution. {{Combining the 6 time-delay distance (6$D_{\Delta t}$) plus 4 angular diameter distance to the deflector (4$D_d$) measurements of time-delay lensed quasars,}} the BAO and the type Ia of supernovae (SNe Ia) datasets, we obtain a 
model-independent result of $r_d = 138.2_{-3.9}^{+3.3}$ Mpc and $H_0 = 72.9^{+1.8}_{-1.8}$ $\Mpc$  for the Taylor series cosmography and $r_d = 137.0_{-3.7}^{+3.2}$ Mpc and $H_0 = 73.1_{-1.7}^{+1.8}$ $\Mpc$ for the Pad\'{e} polynomials cosmography.
The determination of $r_d$ and $H_0$ prefers larger $H_0$ and smaller $r_d$ than Planck data under the assumption of flat-$\Lambda$CDM model. However, the values of $H_0$ are consistent with the $H_0$ determination from SH0ES collaboration. }å

\end{abstract}


\section{Introduction}
Although the standard cosmological scenario, the so-called $\Lambda$-Cold Dark Matter ($\LCDM$) model, provides a remarkable fit to a varies of available cosmological observations, it faces some significant challenges such as tension between the late-time and early-time measurements of the Hubble constant ($H_0$). Assuming that our universe is homogeneous and isotropic at large scales, the SH0ES collaboration \citep{Riess:2021jrx} gives $H_0 = 73.04\pm 1.04 \,\rm{km}\, \rm{s}^{-1}\, \rm{Mpc}^{-1}$ using a distance ladder consisting of Cepheid calibrated Supernova (SN) luminosity distance from Pantheon+ \citep{Scolnic:2021amr}. This value of $H_0$ is in about $5\sigma$ tension with the results from Planck collaboration \citep{Aghanim:2018eyx} using the cosmic microwave background (CMB) radiation measurements assuming a spatially flat standard $\LCDM$ universe, which gives $H_0 = 67.4\pm 0.5 \,\rm{km}\, \rm{s}^{-1}\, \rm{Mpc}^{-1}$.  More discussions on Hubble tension please see the references \citep{Shah:2021onj,2019PhRvL.122v1301P,2019PhRvL.122f1105F,2021ApJ...912..150D,2022A&A...668A..51L,2020ApJ...895L..29L,
2023arXiv230306974D,2023ApJS..264...46L,2022Galax..10...24D} and  references therein for a more comprehensive discussion. However, the Planck's results are based on $\Lambda$CDM model.  Thus, cosmological model-independent measurements for Hubble constant are worth developing, which can potentially reduce tension on measurements of Hubble constant induced by assuming a specific cosmological model.

The Hubble constant is a central parameter in cosmology as it describes the current expansion rate of the universe. The precise measurement of this parameter is crucial for understanding the evolutionary history of the universe and exploring the nature of dark matter and dark energy. 
The discrepancy between early-time and late-time measurements might imply new-physic beyond $\LCDM$ and thus alternative cosmological models have been proposed, including early dark energy models \citep{PhysRevD.87.083009,PhysRevLett.122.221301}, late dark energy models \citep{Zhao:2017cud},  emergent dark energy models \citep{pub.1121080314,Pan_2020,Li:2020ybr,Yang_2021}. On the other hand, there also might be unaccounted for systematic errors or some unknown physical phenomenon that has not been taking into consideration in the current models. For a wide discussion and review about Hubble tension, we refers the readers to \cite{DiValentino:2021izs}.

Therefore, new approaches to determine $H_0$ from local direct measurements were proposed. For example, through time-delay cosmography, the $H_0$ Lenses in COSMOGRAIL's Wellspring (H0LiCOW) Collaboration \citep{10.1093/mnras/stx483} measured the Hubble constant from strong gravitational lens systems with time delays between the multiple images. \cite{Freedman_2021} and \cite{Anand_2022} calibrate the SN Ia with tip of the red giant branch (TRGB) distances to host galaxies, and found $H_0 = 69.8 \pm 0.6$ (stat) $\pm 1.6$ (sys) $\rm{km}\, \rm{s}^{-1}\, \rm{Mpc}^{-1}$ and $H_0  = 71.5 \pm 1.8$ $\rm{km}\, \rm{s}^{-1}\, \rm{Mpc}^{-1}$ respectively, both consistent to within $H_0$ of the SH0ES result. Using detected gravitational waves (GW) as a ``standard siren", binary neutron-star (BNS) system detection GW170817 and subsequent observations in the electromagnetic (EM) domain provide another independent method of measuring the Hubble constant \citep{LIGOScientific:2017adf}.

Additionally, the value of $H_0$ derived from the Planck CMB data is intrinsically linked to the sound horizon at the epoch of last scattering, which is closely associated with the sound horizon $r_d$ at baryon decoupling. Consequently, several modifications to the $\LCDM$ model have been proposed with the objective of alleviating the Hubble tension by decreasing $r_d$ and increasing $H_0$ \citep{Karwal:2016vyq,Anchordoqui:2019yzc,Niedermann:2020dwg,Archidiacono:2020yey}. Nonetheless, \cite{Pogosian:2020ded,Jedamzik:2020zmd} have demonstrated that any model capable of reducing $r_d$ cannot completely resolve the Hubble tension. In an effort to disentangle the degeneracy between $r_d$ and $H_0$, numerous endeavors have been undertaken to constrain $H_0$ and $r_d$ by combining BAO measurements with SNe Ia and observational Hubble data (OHD) \citep{Cai:2022dkh,Zhang:2023eup}. In these studies, while the Hubble tension can be somewhat alleviated, a comprehensive resolution remains absent.

In this research, we plan to jointly constrain $H_0$ and $r_d$ using the latest late-time cosmological observational data including BAO, gravitational lensing time delay and SNe Ia. Specially, we use the newly released BAO measurements from DESI \citep{DESI:2024mwx}. Cosmological results from the measurements of BAO give $H_0 = 68.52 \pm 0.62 \, \Mpc$ when  combined with a baryon density prior from Big Bang Nucleosynthesis and the robustly measured acoustic angular scale from the CMB. For the gravitational lensing time delay, the sample of 6 lensed quasars from the H0LiCOW collaboration is adopted and we use SNe Ia from Pantheon sample.
The reason for using this combination of data is that the absolute distance provided by strong gravitational lensing helps anchor the relative distance of BAO, and supernovae provide a robust history of universe evolution. Meanwhile, we adopted two cosmographic methods, the Taylor series \citep{BOSS:2014hhw,DES:2018rjw,DES:2024ywx,2022ApJ...939...37L} and and Pad\'{e} polynomials \citep{Gruber:2013wua} to investigate $H_0$ and $r_d$ with the dataset including BAO measurements, gravitational lensing time delay and SNe Ia.

This paper is organised as follows: In Section~\ref{sec:cosmo}, we introduced the cosmographic approaches used in this work. In Section~\ref{sec:method}, the methodology and the data we used are presented in detail. And we show our results and discussions in Section~\ref{sec:res}. Finally, the conclusions are drawn in Section~\ref{sec:con}.

\section{Cosmological Approach} \label{sec:cosmo}
There exists a possibility that the $H_0$ discrepancy arises from the cosmological assumption, which cannot be disregarded. Consequently, numerous efforts have been undertaken to avoid the dependence on cosmological models, including approaches such as cosmography, the Parameterization based on cosmic Age (PAge), the Gaussian process (GP) method and so on \citep{10.1093/mnras/stz1163,doi:10.1142/S0218271819300167,pub.1133737479,PhysRevD.106.063519,2024arXiv241114154L}. Among the model-independent methods, cosmography profits from an uncomplicated assumption of homogeneity and isotropy, and can assist us in investigating the evolution of the universe. By using the Friedman-Lemaitre-Robertson-Walker (FLRW) metric, one can express cosmographic parameters by scale factor derivatives with respect to cosmic time as follows \citep{Rezaei:2019xwo}:
\begin{equation} \label{eq:Hz_Taylor}
    H = \frac{{a}^{(1)}}{a}, q = -\frac{{a}^{(2)}}{aH^2}, j =\frac{{a}^{(3)}}{aH^3},s = \frac{{a}^{(4)}}{aH^4},
\end{equation}
where $a^{(n)}$ denotes the $n^{th}$ derivatives of the scale factor $a$. It is clear that these cosmography parameters are independent of cosmological models, and $q_0$, $j_0$ and $s_0$ are the deceleration, jerk, and snap parameters, respectively, which are the second, third, and fourth derivatives of the scale factor with respect to time. We refer the readers to \cite{Mehrabi:2020zau} for more details.
\subsection{Taylor series for cosmology}
We first consider a fourth-order expansion in redshifts following the work in \citet{DES:2018rjw,Visser:2003vq}. The Hubble parameter can be parameterized as
\begin{equation}
\begin{aligned}
    H(z)= &\frac{\dot{a}}{a}=H_0{E(z)}=H_0 [1+(1+q_0)z+(j_0-q_0^2)z^2/2\\&+(3q_0^3+3q_0^2-4q_0j_0-3j_0-s_0)z^3/6],
    \end{aligned}
\end{equation}
then the luminosity distance could be written as
\begin{equation}\label{eq:DL_taylor}
\begin{aligned}
            D_L(z) =& \frac{c(1+z)}{H_0}\int_0^z\frac{dz}{E(z)}\\
         = &\frac{c}{H_0}[ z+(1-q_0)z^2/2-(1+j_0-\Omega_k-q_0-\\&3q_0^2)z^3/6 +d_4z^4/24 ],
\end{aligned}
\end{equation}
where
\begin{equation}
    \begin{aligned}
        d_4 = &2+s_0+5j_0-2\Omega_k-2q_0\\+ &10j_0q_0-6\Omega_kq_0-15q_0^2-15q_0^3,
    \end{aligned}
\end{equation}
and $\Omega_k$ is the curvature. The subscript $0$ signifies the parameters evaluated at the present epoch.

\subsection{Rational Pad\'{e} polynomials for cosmography}

Consider that the various data used in this work involve high redshifts $(z>1)$.
For example, strong lensing systems up to the redshift of $z_s = 1.789$ and SNe Ia  up to the redshift $z=2.3$.  {{Consequently, the convergence of the cosmographic Taylor series may raise significant concerns. A logical methodological alternative may involve the utilization of the Pad\'{e} approximation.}} This method was  firstly proposed  and used for cosmographic analysis by \citet{Gruber:2013wua}, which is much better than Taylor series expansion, as the convergence radius of Pad\'{e} approximation  is larger than Taylor Series expansion.

The Pad\'{e} approximation of an arbitrary function $f (z)$ is given by the rational polynomial
\begin{equation}
    P_{m,n}(z) = \frac{a_0+a_1z+...+a_mz^m}{1+b_1z+...+b_nz^n},
\end{equation}
where the two non-negative integers, $m$ and $n$, are the degrees of the numerator and the denominator, respectively. Considering a Taylor expansion as $f(z)= \sum_{i=0}^\infty c_i z^i$ and equalizing it with the Pad\'{e} approximation, the coefficients $a_i$ ($0\leq i \leq m$) and $b_j$ ($1\leq j \leq n$) are determined by solving the functions:
\begin{equation}
    \sum_{i=0}^\infty c_i z^i = \frac{a_0+a_1z+...+a_mz^m}{1+b_1z+...+b_nz^n}.
\end{equation}

Following the work \citep{Capozziello:2020ctn}, we let the luminosity distance $D_L(z)$ be approximated by the Pad\'{e} approximation and we consider the most stable order of the Pad\'{e} series $m = 2,\, n = 1$ which was demonstrated in \cite{Capozziello:2020ctn} based on the constructions of Pad\'{e} series and on mathematical rules derived from the degeneracy among coefficients.
The Pad\'{e} approximation of the luminosity distance for $m = 2,\, n = 1$ is
\begin{equation}\label{eq:DL_PADE}
    P_{2,1} = \frac{1}{H_0}\left[\frac{z(6(-1+q_0)+(-5-2j_0+q_0(8+3q_0))z)}{ -2(3+z+j_0z)+2q_0(3+z+3q_0z) }  \right].
\end{equation}
The relation between Hubble parameter $H(z)$ and luminosity distance $D_L(z)$ is parameterized as
\begin{equation}
    H(z) = \left[ \frac{\text{d}}{\text{d} z} \left(\frac{ D_L(z)    }{1+z} \right ) \right]^{-1}.
\end{equation}

\vspace{0,5cm}
With the definition of $D_L(z)$ in equation~(\ref{eq:DL_taylor}) for the Talor series  and equation~(\ref{eq:DL_PADE}) for the Pad\'{e} approximation, one can calculate the angular diameter distance through
\begin{equation}\label{eq:Hz_PADE}
    D_A(z) = D_L(z)/(1+z)^2.
\end{equation}
Overall, both of the approaches to joint modeling the expansion history during our analysis are independent of cosmological models.

\section{Methodology and Data} \label{sec:method}

\subsection{Baryon Acoustic Oscillations}
BAO refer to regular, periodic fluctuations in the density of visible baryonic matter in the Universe (the LSS). Being ``the statistical standard ruler", they are commonly used to perform cosmological constraints. In this work, we use measurements of BAO provided by the DESI collaboration \citep{DESI:2024mwx}. The DESI-BAO provided twelve measurements in seven redshift bins from more than 6 million extragalactic objects in the redshift range $0.1 < z < 4.2$.

The BAO measurements used in this work were summarized in Table 1 of \cite{DESI:2024mwx}.
As mentioned in their work, for the Bright Galaxy Sample (BGS, $0.1 < z < 0.4$) and Quasar Sample (QSO, $0.8 < z < 2.1$) tracers, only the angle-averaged $D_V/R_d$ quantity was measured, due to the lower signal-to-noise achieved. For other tracers, including the Luminous Red Galaxy Sample (LRG, 0.4 < z < 0.6 and 0.6 < z < 0.8), the Emission Line Galaxy Sample (ELG, $1.1 < z < 1.6$) and the Lyman$-\alpha$ Forest Sample (Ly$\alpha$, $1.77 < z < 4.16$), the results of $D_M/r_d$, $D_H/r_d$ and the value of the correlation between them, $r$, were given. Where $r_d \equiv r_s(z_*)$ is the physical scale set by the sound horizon $r_s$ at the end of the drag epoch. The distances $D_H(z)$ and $D_M$ are the Hubble and transverse comoving distances respectively and can be obtained with
\begin{equation}
    D_H(z) = \frac{c}{H(z)},
\end{equation}
and
\begin{equation}
    D_M(z) = \frac{D_L(z)}{1+z},
\end{equation}
where $c$ is the speed of light, with $H(z)$ defined in equation~(\ref{eq:Hz_Taylor}) for the Taylor series and in equation~(\ref{eq:Hz_PADE}) for the Pad\'{e} approximation. Similarly, $D_L(z)$ is outlined in equation~(\ref{eq:DL_taylor}) for the Taylor series and in equation~(\ref{eq:DL_PADE}) for the Pad\'{e} approximation. The dialation scale, $D_V(z)$ is a combination of the two distances and defined as
\begin{equation}
    D_V(z)\equiv \left[zD_M^2(z)D_H(z)   \right]^{1/3}.
\end{equation}

Lastly, we treat the value of $r_d$ as a free parameter during our analysis. To make constraints from DESI-BAO measurements, the $\chi^2$ are calculated through
\begin{equation}
    \chi^2_{\text{BAO}} =\Delta ^T(Cov)^{-1}\Delta,
\end{equation}
where $\Delta$ is the difference between the measurements and the associated values determined with the cosmological approach and the $Cov^{-1}$ is the inverse covariance matrix for the BAO measurements.

\subsection{Time-Delay Lensing}
In strong lensing system, the light from background sources is deflected by the intervening mass to the extent that multiple images arise. If the source is variable, the light-paths through different images result in a measurable time-delay between the observed lightcurves \citep{1964MNRAS.128..307R}, $\Delta t = D_{\Delta t} \Delta \Phi (\xi_{\text{lens}}) $, where $\Delta \Phi$ is the Fermat potential difference between two images, which is a function of lens mass profile parameters $\xi_{\text{lens}}$, determined by high-resolution imaging of the host arcs. $D_{\Delta t}$ is the time-delay distance
\begin{equation}
    D_{\Delta t} = (1+z_d)\frac{D_dD_s}{D_{ds}},
\end{equation}
which is a combination of three angular diameter distance $D_d$, $D_s$ and $D_{ds}$ where the subscripts $d$ denote the deflector (lens) and $s$ denote the source. The time-delay distance measurements can be used as a one-rung distance ladder and are independent of the Cepheid distance ladder and early Universe physics. Moreover, the angular diameter distance to the deflector, $D_d$, can be obtained independently of the time-delay measurements and thus provides additional constraints on the expansion history beyond the time-delay measurements.

The latest sample of strong-lensing systems with time-delay observations, recently released by the H0LiCOW Collaboration, consists of six lensed quasars covering the redshift range 0.654 < zs < 1.789 \citep{H0LiCOW:2019pvv}: B1608+656 \citep{2010ApJ...711..201S,doi:10.1126/science.aat7371}, RXJ1131-1231 \citep{Suyu:2012aa,Suyu:2013kha}, HE0435-1223 \citep{H0LiCOW:2016qrm,H0LiCOW:2019xdh}, SDSS1206+4332 \citep{H0LiCOW:2018tyj}, WFI2033-4723 \citep{H0LiCOW:2019mdu}, and PG1115+080 \citep{H0LiCOW:2019xdh}. The observables including $z_d$, $z_s$, $D_d$ and $D_{\Delta t}$ for these lensed quasar systems are summarized in Table 2 of \citet{H0LiCOW:2019pvv}. We note for the lens of B1608+656 that its $D_{\Delta t}$ measurement is given in the form of a skewed log-normal distribution (due to the absence of blind analysis of relevant cosmological quantities), while the derived $D_{\Delta t}$ for the other five lenses are given in the form of Markov chain Monte Carlo (MCMC) distributions. For the measurements of the angular diameter distance to the lens $D_d$, only four strong lensing systems (B1608+656, RXJ1131-1231, SDSS1206+4332, and PG1115+080) are used in our statistical analysis, which are provided in the form of MCMC distributions. We note here that a kernel density estimator is used to compute the posterior distributions of $\mathcal{L}(D_{\Delta t},D_d)$ or $\mathcal{L}(D_d)$ from chains, which accounts for any correlations between $D_{\Delta t}$  and $D_d$ in $\mathcal{L}(D_{\Delta t},D_d)$. The posterior distributions for the six time-delay distances (denoted $6D_{\Delta t}$ for simplicity) and four angular diameter distances to the lenses (denoted $4D_d$) are available at the H0LiCOW website \footnote{http://www.h0licow.org}. For more work on cosmology using strong lensing time delays, we refer the reader to the literature \citep{Kumar:2014vvy,Liao:2019qoc,Liao:2020zko,Bag:2021qnx,Ding:2020jmg}.

In the end, we can write the $\chi^2$ from the time-delay measurement as
\begin{equation}
 \chi^2_{{D_{\Delta t}+D_d}} = -2 \ln \mathcal{L}(D_{\Delta t},D_d) ,
\end{equation}
and
\begin{equation}
 \chi^2_{{D_d}} = -2 \ln  \mathcal{L}(D_d).
\end{equation}
\subsection{Supernavae Ia}

In their role as standard candles, SNe Ia have been of great importance to measure cosmological distances. In our analysis, we use the new ``Pantheon' sample. Previously, \citet{2018ApJ...859..101S} combined the subset of 279 Pan-STARRS1(PS1) ($0.03 < z < 0.68$) supernovae \citep{2014ApJ...795...44R,2014ApJ...795...45S} with the useful data of SNe Ia from SDSS, SNLS, and various low redshift and Hubble Space Telescope (HST)  samples to form the largest combined sample of SNe Ia consisting of a total of 1048 SNe Ia ranging from $0.01<z<2.3$, which is known as the  ``Pantheon" sample. We refer to the work \citep{2018ApJ...859..101S} for more details about the SNe Ia standardization process including the improvements of the PS1 SNe photometry, astrometry and calibration.
\begin{table*}\renewcommand\arraystretch{1.8}
    \caption{Constrain results determined with the combination of different observations. These results are the medians of the marginalised posterior with $68.3\%$ integrated uncertainties.}
    \centering
    \renewcommand{\arraystretch}{1.8}
    \begin{tabular}{cc|ccccc}
    \hline
         Cosmological model  &  Data &  $H_0\,\,[{\Mpc}]$ & $r_d\,\,[{\rm{Mpc}}]$  & $q_0$  & $j_0$ & $s_0$  \\
         \hline
         & DESI+4$D_d$ &  $81.0^{+6.5}_{-7.5}       $ & $126.1^{+9.9}_{-12.0} $ &$-0.465^{+0.027}_{-0.024} $ &   $0.618^{+0.080}_{-0.080} $ & $-0.434^{+0.065}_{- 0.065}$    \\
         Taylor series & DESI+4$D_d$+6$D_{\Delta t}$  &$72.9^{+1.9}_{-1.9}    $   &$139.0^{+ 4.3}_{-4.3}   $  &$-0.469^{+0.025}_{-0.025}$    &$0.639^{+0.078}_{-0.078} $    &$-0.422^{+ 0.065}_{-0.065}$ \\
         & DESI+4$D_d$+6$D_{\Delta t}$+SN &$72.9^{+ 1.8}_{- 1.8}$ &$138.2^{+3.3}_{-3.9}$ &$-0.460^{+ 0.013}_{-0.013} $&$0.656^{+0.065}_{-0.057}$ & $-0.437^{+ 0.050}_{-0.050}$  \\
         \hline
         & DESI+4$D_d$ &$84.2^{+7.0}_{-8.4}$ &$126.4^{+9.6}_{-12.0}$ &$-0.87^{+0.32}_{-0.22}$ &$3.7^{+1.1}_{-2.5}$ &- \\
         Pad\'{e} polynomials & DESI+4$D_d$+6$D_{\Delta t}$  &$ 76.5^{+2.6}_{-2.6}$  &$ 139.3^{+4.2}_{-4.2}$  &$ -0.93^{+0.29}_{-0.25}$  &$ 4.1^{+1.3}_{-2.5}$&-  \\
         & DESI+4$D_d$+6$D_{\Delta t}$+SN & $73.1^{+1.8}_{-1.7}$ & $137.0^{+3.2}_{-3.7}$ & $-0.505^{+0.024}_{-0.024}$ & $1.32^{+0.15}_{-0.15}$ &-  \\
         \hline
    \end{tabular}
    \label{tab:bf-values}
\end{table*}

\begin{table*}\renewcommand\arraystretch{1.5}
    \caption{DIC values for the Taylor series cosmography and Rational Pad\'{e} polynomials.}
    \centering\renewcommand\arraystretch{1.5}
    \begin{tabular}{c|ccc}
    \hline
           Data &   DESI+4$D_d$ & DESI+4$D_d$+6$D_{\Delta t}$ & DESI+4$D_d$+6$D_{\Delta t}$+SN    \\
         \hline
         Taylor series        &  71.70 & 155.10 & 1190.87 \\
         Pad\'{e} polynomials &  46.57 & 138.28 & 1194.58 \\
         $\Delta $DIC         &  25.13 & 16.82  &-3.71  \\

         \hline
    \end{tabular}
    \label{tab:my_label}
\end{table*}

For each SN Ia, the observed distance module is given by
\begin{eqnarray}
\mu_{\mathrm{SN}}=m_B-M_B,\label{con:E2.1}
\end{eqnarray}
where $m_B$ is the observed magnitude in the rest-frame \textit{B}-band. The theoretical  distance modulus $\mu_{\mathrm{th}}$ is defined as
\begin{eqnarray}
\mu_{\mathrm{th}}=5\log_{10}\left[\frac{D_L(z)}{\mathrm{Mpc}}\right]+25,\label{con:E2.2}
\end{eqnarray}
where ${D_L(z)}$ is the luminosity distance associated with the cosmological parameters. Constraining cosmological parameters is implemented by minimizing the $\chi^2$ function:
\begin{equation} \label{Eq:chi2}
\chi^2_{SN}=\sum_{i=1}^{1048}\frac{[\mu_{\mathrm{SN}}(z_i)-\mu_{\mathrm{th}}(z_i)]^2}{\sigma^{2}_{\mu_{\mathrm{SN}}(z_i)}}.
\end{equation}
The nuisance parameter $M_B$ is marginalized.

\section{Results and Discussions} \label{sec:res}

   \begin{figure*}
   \centering
   \includegraphics[width=0.80\textwidth]{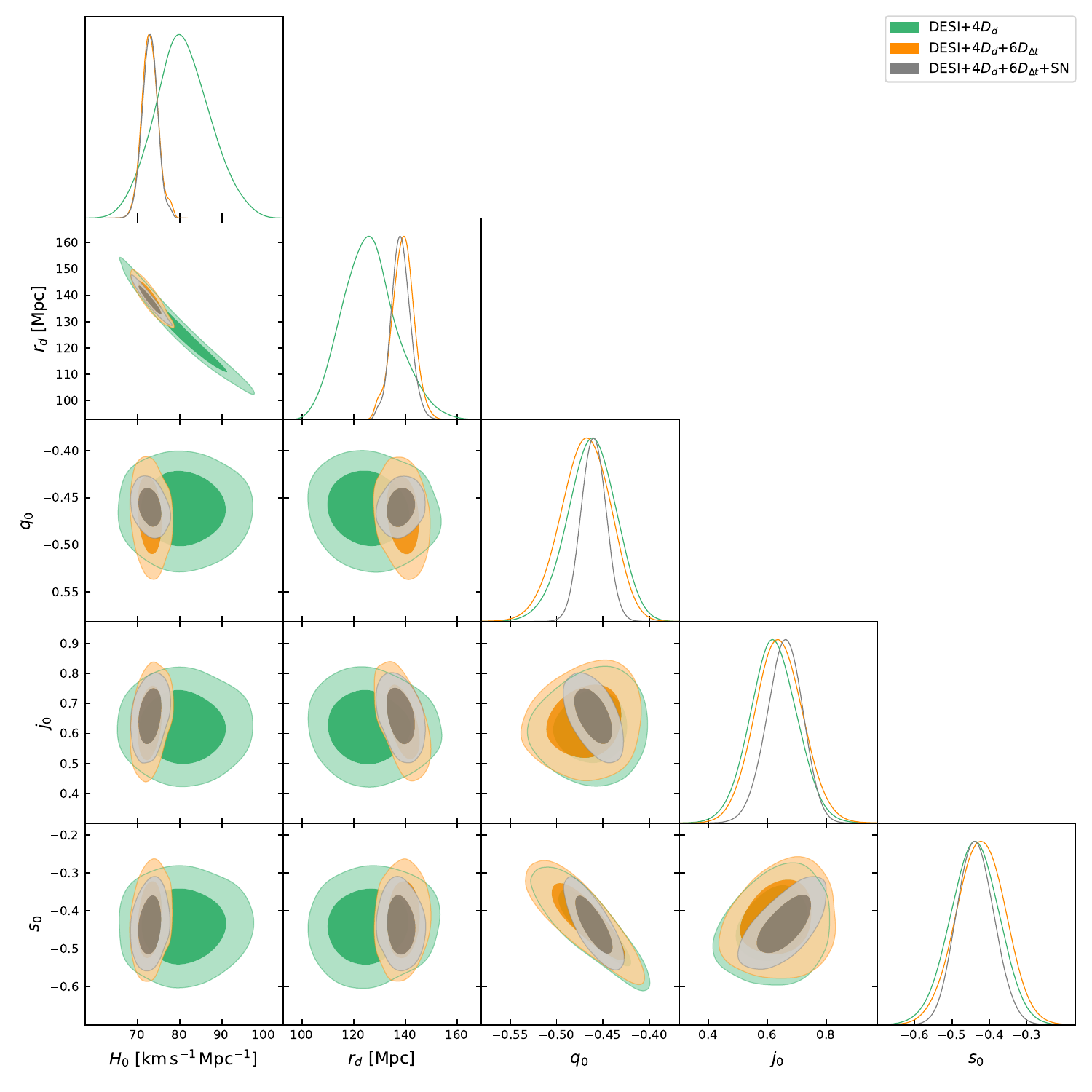}
   \caption{The constraints on $H_0$, $r_d$ and the Taylor expansion cosmographic parameters using the combined DESI-BAO, Time-Delay Lensing and SN Ia. Contours represent the $68.3\%$ and $95.5\%$ confidence intervals. We show the results from different data combinations in different colors.}
              \label{Fig:tri}%
    \end{figure*}
    \begin{figure*}
   \centering
   \includegraphics[width=0.80\textwidth]{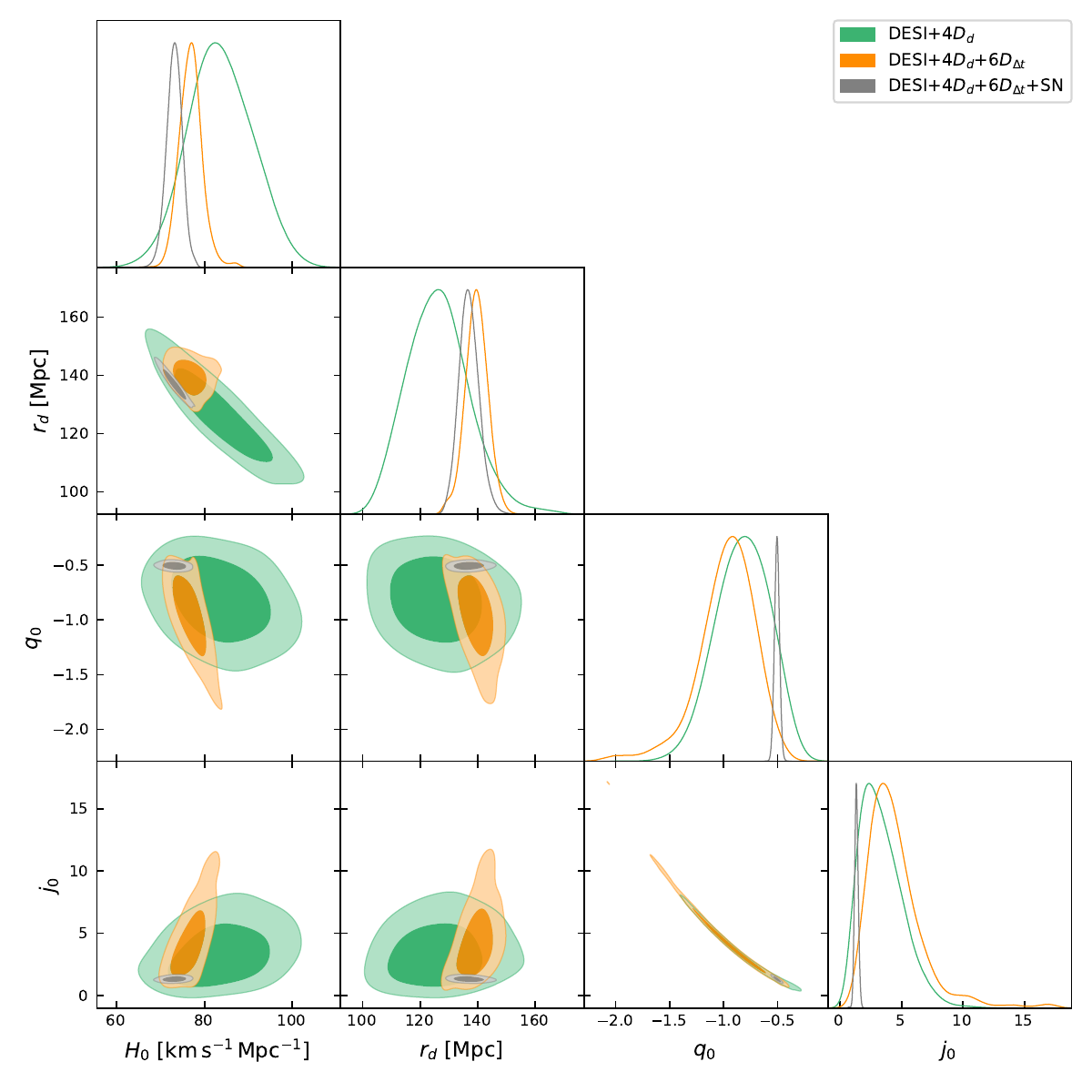}
   \caption{Same as Figure~\ref{Fig:tri} but for the Pad\'{e} polynomials model.}
              \label{Fig:triPADE}%
    \end{figure*}

    \begin{figure*}
   \centering
   \includegraphics[width=0.8\textwidth]{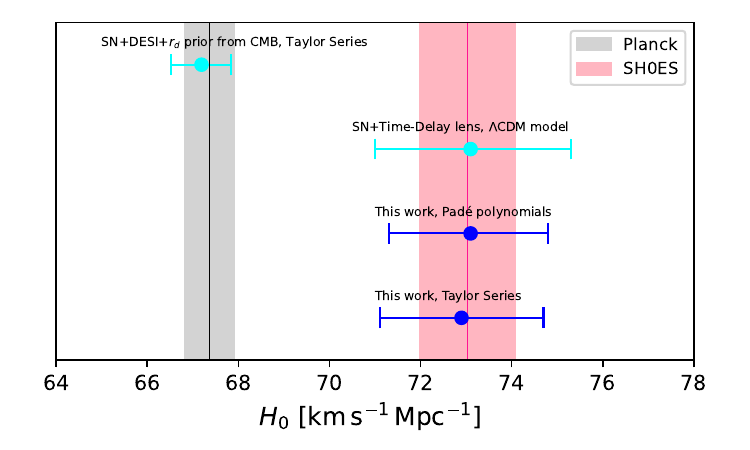}
   \caption{A comparison of Hubble constant values from different data combinations.}
              \label{Fig:H0values}%
    \end{figure*}
In summary, we consider different combinations of time-delay measurements, BAO and SN Ia measurements by adding the $\chi^2$, which is given by following form:
 \begin{equation}
     \chi^2_{All} = \chi^2_{SN}+\chi^2_{\text{BAO}}+ \chi^2_{{D_{\Delta t}+D_d}},
 \end{equation}
if the data is considered.

Based on the method described above, we use a Python package named emcee \citep{Foreman-Mackey:2012any} to do the Markov Chain Monte Carlo analysis, and flat priors are used for each parameter. We analyze the final results with Getdist \citep{Lewis:2019xzd}.

The best-fitting parameters numerical results determined with the combination of the SN Ia, DESI-BAO and time-delay lensing datasets are shown in Table~\ref{tab:bf-values} for two cosmographical  methods. These results are the medians of the marginalized posterior with $68.27\%$.  The posterior joint distributions along with  different data combinations and cosmographical methods are shown in Figure~\ref{Fig:tri} and Figure~\ref{Fig:triPADE}, respectively.
Let's first focus on the  constraining  result of $H_0$.
The combined DESI-BAO, Time-Delay Lensing and SN Ia datasets yield $H_0 = 72.9_{-1.8}^{+1.8}$ $\text{km} \,\, \text{s} ^{-1} \,\,\text{Mpc} ^{-1} $ and $H_0 = 73.1^{+1.8}_{-1.7}$ $\text{km} \,\, \text{s} ^{-1} \,\,\text{Mpc} ^{-1} $ (median of the marginalized posterior with $68.27\%$ integrated uncertainties) when using the Taylor series cosmography and the Pad\'{e} polynomials cosmography, respectively. These results are consistent with the SH0ES collaboration measurements of $H_0 = 73.04_{-1.04}^{+1.04}$ $\text{km} \,\, \text{s} ^{-1} \,\,\text{Mpc} ^{-1} $  \citep{Riess:2021jrx} and previous constrain results from SN Ia in combination with Time-Delay lensing assuming flat-$\Lambda$CDM model, which gives  $H_0 = 73.1^{+2.1}_{-2.2}$ $\text{km} \,\, \text{s} ^{-1} \,\,\text{Mpc} ^{-1} $  \citep{Taubenberger:2019qna}. However, these measurements are in tension with the Planck+Flat-$\Lambda$CDM measurement of $H_0 = 67.4 _{-0.5}^{+0.5} \,\text{km} \,\, \text{s} ^{-1} \,\,\text{Mpc} ^{-1} $ and the inverse distance ladder measurements from the combined DES-SN5YR + DESI-BAO datasets that yield $H_0 = 67.19_{-0.66}^{+0.66} \,\text{km} \,\, \text{s} ^{-1} \,\,\text{Mpc} ^{-1}$ \citep{DES:2024ywx}. Compared with results between the combinations of  DESI+4$D_d$+6$D_{\Delta t}$ and DESI+4$D_d$+6$D_{\Delta t}+SN$, we find that the SN data did not contribute much to the constraining chances of $H_0$ and $r_d$, and more to improving the constrains of the cosmographic parameters ($q_0$, $j_0$ and $s_0$). This is perfectly consistent with our motivation in the introduction, i.e., Time-Delay Lensing data help anchor relative distances and contribute to constraining results for $H_0$ and $r_d$, SN data provide a robust history of universe evolution and contribute to cosmographic parameters.
For combination of DESI+4$D_d$, the results for Hubble constant  are $81.0^{+6.5}_{-7.5}$ $\Mpc$ and  $84.2^{+7.0}_{-8.4}$ $\Mpc$. This result is reasonable. In the results of H0LiCOW analyzed using only $D_d$ for measuring $H_0$, their results demonstrate a precision of about 10\% level for $H_0$. As demonstrated in the results in the work of \citet{2019Sci...365.1134J}, they used two gravitational lenses to determine the angular diameter distances and provided a result of $H_0=82.4^{+8.4}_{-8.3}$ $\Mpc$.  This is because the inference of angular diameter distances requires additional information  (the spectroscopic measurement of the stellar kinematics of the lens galaxy) compared to time delay distances, more observational errors and systematic uncertainties are introduced for inferencing $D_d$. Our results are consistent with the results reported by the work \citep{2019Sci...365.1134J}.
In Figure~\ref{Fig:H0values} we show a direct comparison between our result for $H_0$ and other results in the literature.

Now let us remark on the sound horizon $r_d$ measurements from the different data combinations.
The final result obtained combining all data gives  $r_d=138.2^{+3.3}_{-3.9}$ Mpc in the framework of the Taylor series method. This result changes to  $r_d=137.0^{+3.2}_{-3.7}$ Mpc when the Pad\'{e} polynomials is used.
It should be emphasized here that there is a strong degeneracy and negative correlation between $r_d$ and $H_0$. The Figure~\ref{Fig:tri} and Figure~\ref{Fig:triPADE}  illustrate this correlation well, and it is possible to show the contribution of various data to the  sound horizon parameter, each with a different role and contribution. It is interesting to compare our results with previous works.  {For example, \citet{2019MNRAS.486.5046W} used the $D_d$ of three lenses from the H0LiCOW collaboration, combined relative distances from SNe Ia and BAOs plus a prior of $H_0$ from H0LiCOW,  they reported a $r_d=137\pm4.5$ Mpc by using cosmography and global fitting method. When their didn't take into account the prior of $H_0$, their reported $H_0=72.3\pm6.9$ $\Mpc$ and $r_d=139.2\pm13.3$ Mpc.  However, their work used BAO data from the the Sloan Digital Sky Survey (SDSS). Our results are almost similar to theirs, suggesting that the BAO data from DESI and SDSS  should be compatible. In addition, our work does not need to consider the prior of $H_0$, and the constraining precision of $H_0$ and $r_d$ is greatly improved compared with previous work.}

In particular, the cosmographic coefficients do not differ much between these two cases by using DESI-BAO, Time-Delay Lensing and SN Ia datasets. The deceleration parameters $q_0=-0.460^{+0.013}_{-0.013}$ and $q_0=-0.505^{+0.024}_{-0.024}$  corresponding to the Taylor series and Pad\'{e} polynomials, respectively, are mutually consistent. Their values indicate an accelerating expansion of the Universe. To better understand the meaning of these parameters, let us consider a flat $\Lambda$CDM model. Then, one can relate cosmographic parameters to the physical quantity $\Omega_{m0}$ (i.e. the matter density parameter at present time) according to $q_0=-1+3/2\Omega_{m0}$. It is easy to check that our results are compatible with a flat $\Lambda$CDM model with the current matter density parameter $\Omega_{m0}=0.31$ obtained from \textit{Planck} CMB observations within 1$\sigma$ confidence level \citep{Aghanim:2018eyx}.

For a quantitative comparison between the higher order cosmographic models, we employ the deviance information criterion (DIC) \citep{10.1111/1467-9868.00353,10.1111/j.1745-3933.2007.00306.x}, which is defined as 
\begin{equation}
    \text{DIC} \equiv  D(\overline{\theta})+2p_D = \overline{D(\theta)}+p_D,
\end{equation}
where $P_D = \overline{D(\theta)}-D(\overline{\theta})$ and $D(\overline{\theta})=-2 \ln{\mathcal{L}} + {C}$, $C$ is a "standardizing" constant deending only on the data that will vanish from any derived quantity, and $D$ is the deviance of the likelihood. If we define an effective $\chi^2$ as usual by $\chi^2 =-\ln {\mathcal{L}}$, we can write
\begin{equation*}
    p_D = \overline{\chi^2(\theta)} -\chi^2(\overline{\theta})
\end{equation*}

We find strong evidence against Taylor series cosmography with $\Delta \text{DIC} = 25.13$ and $\Delta \text{DIC} = 16.82$ for DESI-BAO in combination with 4$D_d$ measurements and DESI-BAO in combination with 4$D_d$+6$D_{\Delta t}$ measurements, respectively. However, the Taylor series cosmography behaves better when adding SN observations to the data combination according to the DIC criterion. There is weak evidence against the Rational Pad\'{e} polynomials model with $\Delta \text{DIC} = -3.17$ for adding SN Ia to DESI-BAO and 4$D_d$+6$D_{\Delta t}$. {{In all,  the DIC results ultimately show that there is no clear cosmological approach preference for different combinations of data, we also take into account the problem of Taylor expansion ($z$ much larger than 1) failures at high redshifts. Therefore, we recommend using Padé approximation and discarding Taylor expansion in future  analyses.}}

\section{Conclusion} \label{sec:con}
In this work, we determine the Hubble constant $H_0$,  sound horizon $r_d$ and other cosmological parameter such as deceleration parameter from observations including recent DESI-BAO, Time-Delay lensing and SN Ia without assuming particular cosmogical model. These data have different contributions,  the absolute distance provided by  Time-Delay lensing  helps anchor the relative distance of BAO, and SN Ia provide a robust history of universe evolution. Considering that redshift coverage of the data we used extends to $z \sim 2.3$, the issue of convergence regarding the Taylor series cosmographic expansion becomes important. Therefore we also used the technique of Pad\'{e} polynomials cosmography.

The combined DESI-BAO, Time-Delay Lensing and SN Ia datasets yield $H_0 = 72.9^{+1.8}_{-1.8}$ $\text{km} \,\, \text{s} ^{-1} \,\,\text{Mpc} ^{-1} $ and $H_0 = 73.1^{+1.8}_{-1.7}$ $\text{km} \,\, \text{s} ^{-1} \,\,\text{Mpc} ^{-1} $  when using the Taylor series cosmography and the Pad\'{e} polynomials cosmography, respectively. These results are consistent with the SH0ES collaboration measurements. 
For constrained results on sound horizon, the final result obtained combining all data gives  $r_d=138.2^{+3.3}_{-3.9}$ Mpc in the framework of the Taylor series method. This result changes to  $r_d=137.0^{+3.2}_{-3.7}$ Mpc when the Pad\'{e} polynomials is used. {{Compared with previous work using SDSS BAO data, our results are almost similar to theirs, suggesting that the BAO data from DESI and SDSS should be compatible. However, it is important to emphasize that DESI is conducting a five-year survey covering 14,200 square degrees of the sky within the redshift range 0.1 < z < 4.2. The spectroscopic sample size for DESI is expected to be an order of magnitude larger than that of previous SDSS surveys. While there is some overlap in the sky regions and redshift ranges observed by DESI and SDSS, the input catalogs used for BAO analyses differ due to the distinct instrument performance characteristics and observing strategies employed by the two surveys. Moreover, future DESI data releases will provide a substantially larger effective volume ({$V_{\text{eff}}$}) in certain redshift ranges, thereby improving statistical precision and constraining power.}} In addition,  we find that the Taylor series cosmography behaves a little better according to the DIC for the data combination.

{As a final remark, it remains important to consider that there might be an unexposed systematic error. This motivates the efforts to continue acquiring more and better data to investigate these cosmological tensions further. For the $H_0$ constraint, we do not expect a different $H_0$ constraint from the updated BAO data,  but we expect that current and future surveys like the Dark Energy Survey (DES) \citep{2018MNRAS.481.1041T}, the Hyper SuprimeCam Survey \citep{2017MNRAS.465.2411M}, and the  Legacy Survey of Space and Time (LSST) \citep{2010MNRAS.405.2579O}
will bring hundreds of thousands of lensed quasars in the most optimistic discovery scenarios to help us achieve high-precision $H_0$ constraints.  On the other hand, the DESI survey will  have access to vast amounts of BAO data and be able to bring down cosmological parameters errors, thus to alleviate these cosmological tensions in higher precision and accuracy measurements.}

\section*{Acknowledgments}
This work was supported by the National Natural Science Foundation of China under Grant No. 12203009; the Science Research Project of Hebei Education Department No. BJK2024134; the Chutian Scholars Program in Hubei Province (X2023007); and the Hubei Province Foreign Expert Project (2023DJC040). This work benefits from the high-performance computing clusters at College of Physics, Hebei Normal University.

\bibliography{sample631}{}
\bibliographystyle{aasjournal}

\end{document}